\documentclass[PRL,twocolumn,aps,longbibliography]{revtex4-1}
\usepackage{graphicx}
\usepackage{txfonts}
\usepackage[T1]{fontenc}
\usepackage{epstopdf}
\usepackage{verbatim}
\usepackage{color}

\begin{document}
\title{Experimental observation of Weyl points}
\author{Ling Lu$^{1}$}
\email{linglu@mit.edu}
\author{Zhiyu Wang$^{2}$}
\author{Dexin Ye$^{2}$}
\author{Lixin Ran$^{2}$}
\author{Liang Fu$^{1}$}
\author{John D. Joannopoulos$^{1}$}
\author{Marin Solja\v{c}i\'{c}$^{1}$}
\affiliation{$^{1}$ Department of Physics, Massachusetts Institute of Technology, Cambridge, Massachusetts 02139, USA}
\affiliation{$^{2}$ Laboratory of Applied Research on Electromagnetics, Zhejiang University, Hangzhou, 310027, China}

\begin{abstract}
\end{abstract}

\maketitle

\textbf{
In 1929, Hermann Weyl derived~\cite{Weyl1929} the massless solutions from the Dirac equation -- the relativistic wave equation for electrons.
Neutrinos were thought, for decades, to be Weyl fermions until the discovery of the neutrino mass.
Moreover, it has been suggested that low energy excitations in condensed matter\cite{volovik2009universe,Wan2011Weyl,Burkov2011WeylMultilayer,Xu2011ChernSemimetal,singh2012topological,liu2014weyl,bulmash2014prediction} can be the solutions to the Weyl Hamiltonian. Recently, photons have also been proposed to emerge as Weyl particles inside photonic crystals~\cite{Lu2013Weyl}.
In all cases, two linear dispersion bands in the three-dimensional~(3D) momentum space intersect at a single degenerate point -- the Weyl point.
Remarkably, these Weyl points are monopoles of Berry flux with topological charges defined by the Chern numbers\cite{volovik2009universe,Wan2011Weyl}.
These topological invariants enable materials containing Weyl points to exhibit a wide variety of novel phenomena
including surface Fermi arcs\cite{potter2014quantum}, chiral anomaly\cite{Aji2012ABJ-Weyl}, negative magnetoresistance\cite{son2013chiral}, nonlocal transport\cite{parameswaran2014probing}, quantum anomalous Hall effect\cite{Yang2011QHEWeyl}, unconventional superconductivity\cite{cho2012superconductivity} and others~\cite{turner2013beyond,hosur2013recent}.
Nevertheless, Weyl points are yet to be experimentally observed in nature.
In this work, we report on precisely such an observation in an inversion-breaking 3D double-gyroid photonic crystal without breaking time-reversal symmetry.
}

\begin{figure}[t]
\includegraphics[width=0.4\textwidth]{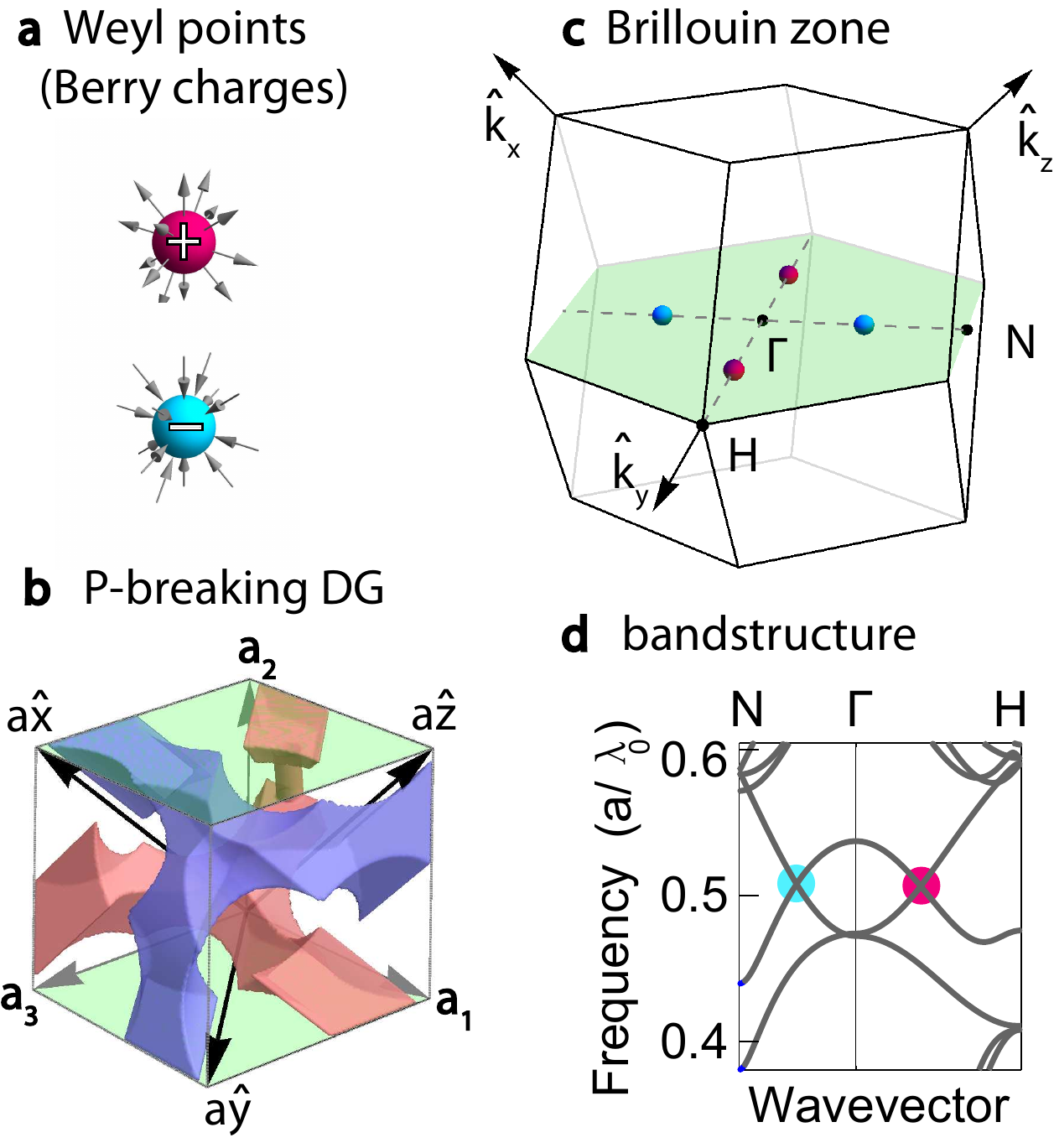}
\caption{Weyl points found in the bandstructure of a $P$-breaking gyroid photonic crystal.
a) The Weyl points are monopoles of Berry flux in the momentum space.
b) The bcc cell of the DG with a $P$-breaking defect on the red gyroid, where
$\mathbf{a_1}=(-1,1,1)\frac{a}{2}$, $\mathbf{a_2}=(1,-1,1)\frac{a}{2}$ and $\mathbf{a_3}=(1,1,-1)\frac{a}{2}$.
The (101) surfaces are highlighted in green.
c) The BZ of the DG photonic crystal in b). Four Weyl points were illustrated on the green (101) plane along $\Gamma-H$ and $\Gamma-N$, where $H=(0,1,0)\frac{2\pi}{a}$ and $H=(-0.5,0,0.5)\frac{2\pi}{a}$.
d) The photon dispersions were plotted along the $N-\Gamma-H$. The Weyl points are the linear band-touchings between the 4th-and 5th bands.
}
\label{Fig:intro}
\end{figure}

\begin{figure}[t]
\includegraphics[width=0.5\textwidth]{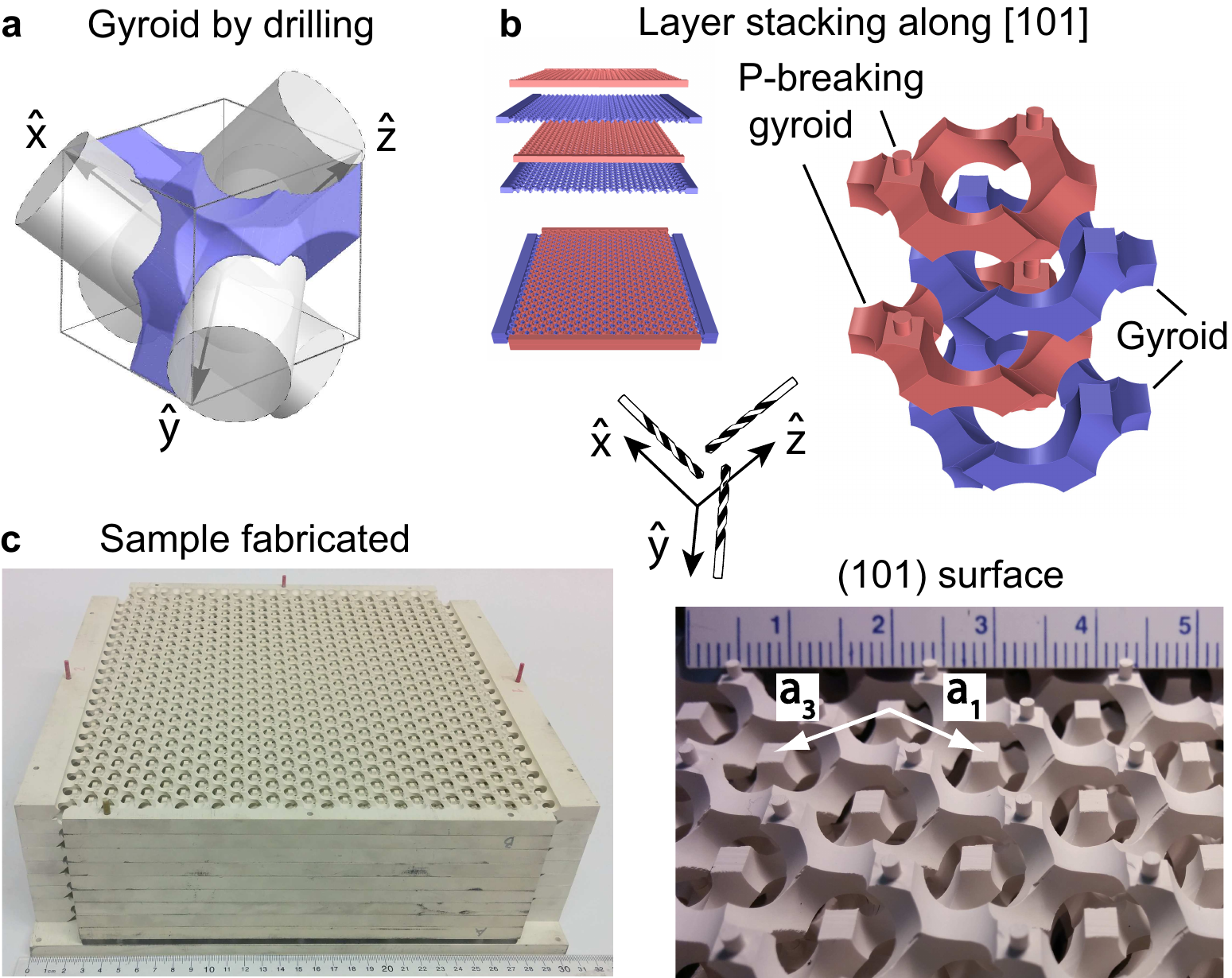}
\caption{Fabrication of gyroids by drilling periodic holes along $x$, $y$, $z$ directions.
a) Illustration in a bcc unit cell that a single gyroid structure can be approximated by drilling air-holes.
b) The double-gyroid structure can be made by stacking layers along the $[101]$ direction. The red and blue gyroids, being inversion counterpart, inter-penetrates each other. We breaking parity~($P$, inversion) by shrinking the vertical connections to thin cylinders for the red gyroid.
c) Shown on the right, a total of 18 layers were stacked. A zoom-in view from top is shown on the left with a ruler~(in centimeters) in the background.
}
\label{Fig:DG}
\end{figure}


Weyl points are sources of quantized Berry flux of $\pm2\pi$ in the momentum space. Their charges can be defined by the corresponding Chern numbers of $\pm1$, as shown in Fig. \ref{Fig:intro}a. So, Weyl points robustly appear in pairs and can only be removed through pair annihilation. Since the Berry curvature is strictly zero under $PT$ symmetry, --- the product of parity~($P$, inversion) and time-reversal symmetry~($T$), isolated Weyl points only exist when at least one of $P$ or $T$ is broken.
In Ref. \cite{Lu2013Weyl}, frequency-isolated Weyl points were predicted in $PT$-breaking DG photonic crystals.
We chose to break $P$ instead of $T$, in the experiment, to avoid using magnetic materials and applying static magnetic fields. This also allows our approach to be directly extended to photonic crystals at optical wavelengths.
This $P$-breaking DG is shown in its body-centered-cubic~(bcc) unit cell in Fig. \ref{Fig:intro}b. At the presence of $T$, there must exist even pairs of Weyl points. The two pairs of Weyl points illustrated in the Brillouin zone~(BZ), in Fig. \ref{Fig:intro}c, are thus the minimum number of Weyl points possible.
The bandstructure plotted in Fig. \ref{Fig:intro}d shows two linear band-crossings along $\Gamma-N$ and $\Gamma-H$. The other two Weyl points have identical dispersions due to $T$.

We work at the microwave frequencies around 10GHz for the accessible fabrication of 3D photonic crystal.
The current additive processes like 3D printing can hardly fulfill the material requirement of low-loss dielectrics with high-dielectric constants.
In order to fabricate the two inter-penetrating gyroids with subtractive processes, we open up each gyroid network by layers along the $[101]$ direction with equal thickness of $\frac{a}{\sqrt{2}}$. The $P$-breaking defects are introduced in each layer of red gyroid in Fig. \ref{Fig:intro}b.

We approximate each gyroid network by three sets of hole-drilling, as illustrated in Fig. \ref{Fig:DG}a in a unit cell of the body-centered-cubic~{bcc} lattice.
Similar methods of drilling and angled etching has been used in fabrication of  3D photonic crystals at microwave\cite{yablonovitch1991photonic} and near infrared wavelengths\cite{Takahashi2009}.
The three cylindrical air holes, of the blue gyroid, along $\hat{x}$, $\hat{y}$ and $\hat{z}$ go through $(0,\frac{1}{4},0)a$, $(0,0,\frac{1}{4})a$ and $(\frac{1}{4},0,0)a$ respectively. All air holes have a diameter of 0.54$a$, where $a$ is the cubic lattice constant. 
Gyroids approximated by this drilling approach have almost identical bandstructures as those defined by the level-set iso-surfaces in Ref. \cite{Lu2013Weyl}.

The second~(red) gyroid is the inversion counterpart of the blue gyroid. To break $P$, we shrink one arm of the red gyroid at $(\frac{1}{4},-\frac{1}{8},\frac{1}{2})a$ to be a cylinder oriented along $[101]$. 
Shown in Fig. \ref{Fig:DG}b, the defect cylinder has a diameter of $0.1a$ and height of $0.2a$. We separate the red gyroid by cutting $(101)$ planes at the centers of the defect cylinders. The blue gyroid, without introduced defects, is separated at the corresponding position of $(-\frac{1}{4},\frac{1}{8},-\frac{1}{2})a$.
Each layer is $\frac{a}{\sqrt{2}}$ thick, shown in Fig. \ref{Fig:DG}b.
The separated layers of each gyroid are identical up to translations of $\mathbf{a_2}$~(the bcc lattice vector).
One unit vertical period of DG consists of two layers from each gyroids, in which each layer is $\frac{a}{\sqrt{2}}$ thick.

\begin{figure*}[t]
\includegraphics[width=\textwidth]{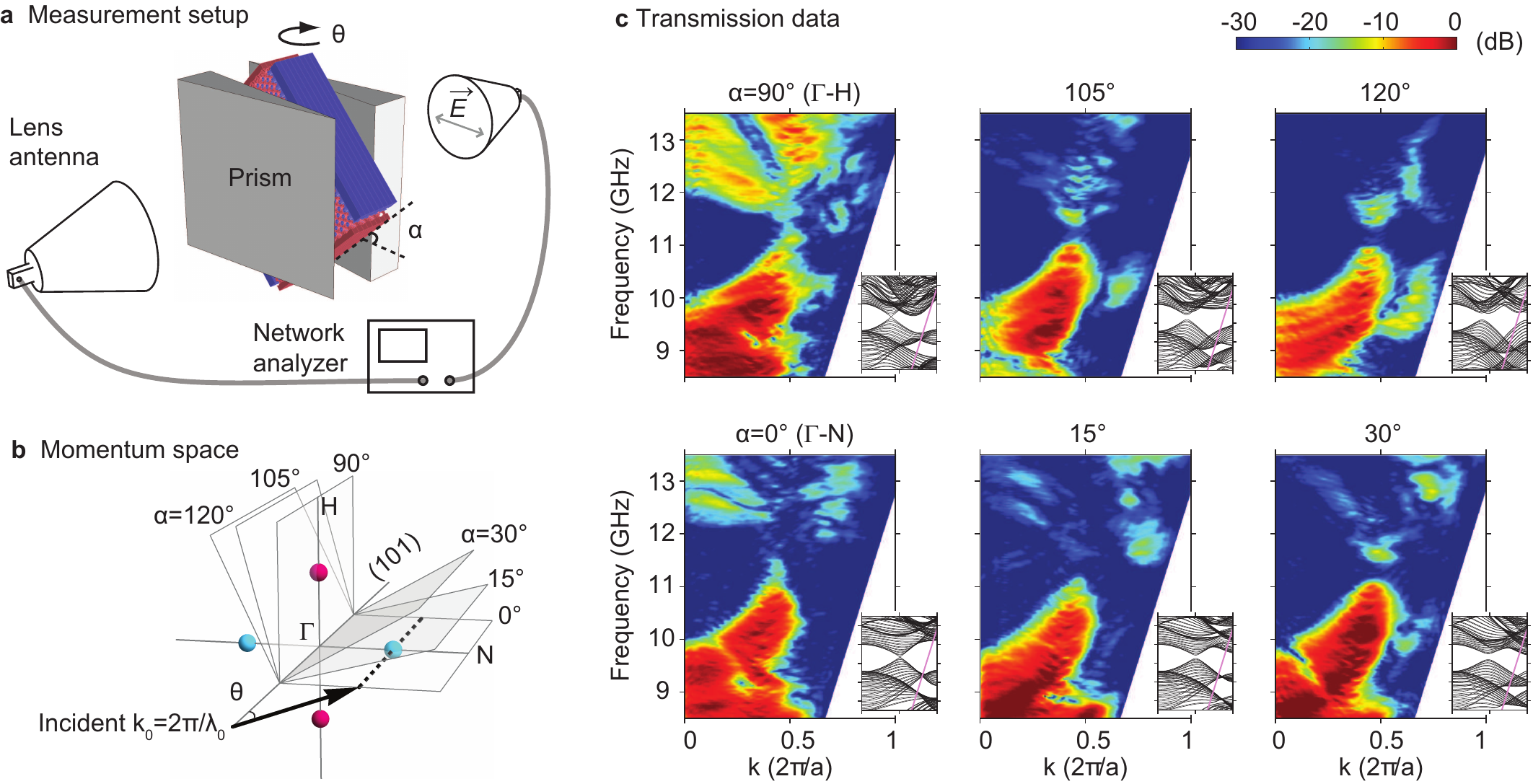}
\caption{Transmission measurement results.
a) Schematic of the microwave transmission setup.
b) The bulk states in the sample are exited by the incident wavevector parallel to the sample surface varied by angle $\theta$.
c) Transmission data as the sample is rotated along the $[101]$ axis by angle $\alpha$. Insets are calculated bandstructures projected along $[101]$; they are scaled to the same range and ratio as the measured data.
}
\label{Fig:Data}
\end{figure*}

The materials of choice are slabs of ceramic-filled plastics~(C-STOCK AK from Cuming Microwave) of dielectric constant of 16 and loss tangent of 0.01. Each slab has a thickness of 9.5$mm$~($=\frac{a}{\sqrt{2}}$).  Both width and length of the slabs are 304$mm$~($\sim$12 inches). The material hardness is adjusted between 80 to 90 on a Shore D gauge to be machined by carbide tools without cracking. Computer-numerical-controlled machining was used to drill the  slabs. Although all slabs of the same gyroid can be machined altogether, we processed no more than two layers at the same time to leave a solid frame after tilted drilling. Each layer experienced around 700 drills along $\pm$45 degrees away from its normal~([101]) and about 40 drills in $\hat{y}$ on both top and bottom sides of the slab.
A frame of $\sim$20mm were left blank on the four sides of the slabs for handling. We subsequently cut two sides of the frames of each gyroid for assembling. Illustrated in Fig. \ref{Fig:DG}b, the layers of the two gyroids were offset by half of the slab thickness~(4.25mm). A picture of the final assembled sample is shown in Fig. \ref{Fig:DG}c. It consists of a total of 18 layers, with 9 layers from each gyroid stacked in alternating order.

We performed angled resolved transmission measurements on the photonic crystal sample to probe the dispersions of the bulk states.
The schematic of the experimental setup is shown in Fig. \ref{Fig:Data}a. A pair of lens antennas
 were placed on the two sides of the sample. Transmission amplitudes~(S21) were recorded by the network analyzer. 
The polarizations of both antennas was horizontal and their half-power beam widths were ~9$^\circ$.
The collimated beam impinged on the sample (101) surface, in which the incident angle is varied by rotating the sample around the vertical axis.
As illustrated in Fig. \ref{Fig:Data}b, the component of the incident wavevectors~[$k_0sin(\theta)$] parallel to the sample surface are conserved up to a reciprocal lattice vector of the sample lattice due to the discreet translational symmetry. All bulk states of the same wavevector projection~(dashed line in Fig. \ref{Fig:Data}b) and frequency could be excited in the bulk and exist the sample in the same direction as the incident beam. The exiting signal was collected by the antenna at the opposite side.

Illustrated in Fig. \ref{Fig:Data}b, the Weyl points in the system locate along $G-H$ and $G-N$ directions about half way between the BZ center~($\Gamma$) and the zone boundaries. To access them from free space, the incident angle $\theta$ would have to be large than 60$^\circ$ and the effective cross-section of the sample would to small for enough signal to go through. To overcome this problem, we placed a pair of angled prisms~(12.4$^\circ$) in contact with the opposite surfaces of the sample as shown in Fig. \ref{Fig:Data}a.
The prisms are made of the same material as the sample~($\epsilon=16$).

We mapped out all bulk states, projected along $[101]$, by varying $\theta$ and $\alpha$ shown in Fig. \ref{Fig:Data}a and b.
Six transmission data of representative directions~($\alpha$) are plotted in Fig. \ref{Fig:Data}c. The figure insets are projected dispersions of the bulk bandstructures along the [101] direction.
The transmission data stops on the right slanted boundary, which corresponds to the maximum rotation angle in $\theta$. Close to this boundary, the transmission intensity is low due to the smaller effective cross-section of the samples at large angles. 

When $\alpha=90^\circ$, the beam scan through the upper Weyl point along $G-H$ represented by a magenta sphere. The data clearly shows a linear point touching at 11.3GHz in frequency and $\frac{\pi}{a}$ in wavevector. As $\alpha$ deviated from 90$^\circ$ to 105$^\circ$ and 120$^\circ$, the point touching opens a gap as expected for a point-degeneracy.
The other Weyl point, in cyan, on the right of the $G-N$ axis was studied by orient $\alpha$ to be 0$^\circ$, 15$^\circ$ and 30$^\circ$. 
Although the transmission intensity of the upper bulk bands is not prominent, the Weyl point dispersions can still be inferred from the curvatures of the lower bulk bands.
In principle, all bulk bands of frequency and momentum matching the incident beam can be coupled and transmitted. However, the coupling and transmission efficiency depend on the details of the Bloch mode polarization, field distribution, group velocities, radiation lifetime and so on. The low transmission amplitude in some parts of the data at high frequency region is due to the mismatch between bulk and the incident waves in free space.
The remaining two Weyl points, at the opposite $\mathbf{k}$ locations, relate to the two measured Weyl points by $T$. They have the same projected band dispersions and same transmission pattern as the data shown in Fig. \ref{Fig:Data}c.
All the transmission data, in Fig. \ref{Fig:Data}c, compares very well with the theoretical bandstructures in the insets except a few regions with low transmission power.

Our experimental demonstration of Weyl points answers the long standing search for a natural realization of the Weyl equations.
These photonic Weyl points pave the way to topological photonics\cite{lu2014topological,Wang2009,hafezi2013imaging,rechtsman2013photonic}
 in 3D, where 3D Dirac points\cite{liu2014discovery} and various gapped topological phases~\cite{Fu2011TCI} can be discovered. Similar approaches can be readily adopted to observe Weyl points at optical frequencies using 3D nanofabrication\cite{turner2011fabrication,crossland2008bicontinuous}. We also anticipate that the Weyl points should be observable in other systems\cite{dubvcek2014weyl}, notably the topological semi-metals\cite{weng2014weyl,huang2015inversion} in condensed matter.

\section*{Acknowledgements}
We thank Andrew Gallant and Ernest Johnson at MIT central machine shop for machining the gyroid layers.
We thank Yichen Shen, Zhang Bin, Junwei Liu, Bo Zhen and Ashvin Vishwanath for discussions.
J.J. was supported in part by the U.S.A.R.O. through the ISN, under Contract No. W911NF-13-D-0001.
L.F. was supported by the DOE Office of Basic Energy Sciences, Division of Materials Sciences and Engineering under Award No. DE-SC0010526.
L.L. was supported in part by the MRSEC Program of the NSF under Award No. DMR-1419807.
M.S. and L.L. were supported in part by the MIT S3TEC EFRC of DOE under Grant No. DE-SC0001299.
\bibliography{Ling}

\end{document}